\documentclass[10pt,letter,prl,twocolumn,hypertext]{revtex4-1}
\usepackage{ifthen}
\newboolean{uprightparticles}
\setboolean{uprightparticles}{false} %Set to true to get roman particle symbols

\usepackage{amssymb}
\usepackage{amsfonts}
\usepackage{upgreek}
\usepackage{xspace}
\usepackage{color}
\usepackage{amsmath}
\usepackage{graphicx}
\usepackage{bm}
\usepackage{dcolumn}% Align table columns on decimal point

\newboolean{articletitles}
\setboolean{articletitles}{true}

\newboolean{pdflatex}
\setboolean{pdflatex}{true} % use this if using non-eps figures

\def\ux85 {UX85\xspace}

\ifthenelse{\boolean{uprightparticles}}%
{

 \def\Ppi         {\ensuremath{\uppi}\xspace}

 \def\PDelta      {\ensuremath{\Delta}\xspace}                 
 \def\PXi      {\ensuremath{\Xi}\xspace}                 
 \def\PLambda      {\ensuremath{\Lambda}\xspace}                 
 \def\PSigma      {\ensuremath{\Sigma}\xspace}                 
 \def\POmega      {\ensuremath{\Omega}\xspace}                 
 \def\PUpsilon      {\ensuremath{\Upsilon}\xspace}                 
 
 \def\PB      {\ensuremath{\mathrm{B}}\xspace}                 
                  
 \def\PD      {\ensuremath{\mathrm{D}}\xspace}

 \def\PK      {\ensuremath{\mathrm{K}}\xspace}

 \def\Pi      {\ensuremath{\mathrm{i}}\xspace}

}
{

 \def\Ppi         {\ensuremath{\pi}\xspace}

 \mathchardef\PDelta="7101
 \mathchardef\PXi="7104
 \mathchardef\PLambda="7103
 \mathchardef\PSigma="7106
 \mathchardef\POmega="710A
 \mathchardef\PUpsilon="7107
                  
 \def\PB      {\ensuremath{B}\xspace}                 
                  
 \def\PD      {\ensuremath{D}\xspace}

 \def\PK      {\ensuremath{K}\xspace}

 \def\Pi      {\ensuremath{i}\xspace}

}

\def\pion  {\ensuremath{\Ppi}\xspace}

\def\pip   {\ensuremath{\pion^+}\xspace}
\def\pim   {\ensuremath{\pion^-}\xspace}

\def\kaon  {\ensuremath{\PK}\xspace}
  \def\Kbar  {\kern 0.2em\overline{\kern -0.2em \PK}{}\xspace}

\def\Kz    {\ensuremath{\kaon^0}\xspace}
\def\Kzb   {\ensuremath{\Kbar^0}\xspace}
\def\KzKzb {\ensuremath{\Kz \kern -0.16em \Kzb}\xspace}
\def\Kp    {\ensuremath{\kaon^+}\xspace}
\def\Km    {\ensuremath{\kaon^-}\xspace}

\def\KpKm  {\ensuremath{\Kp \kern -0.16em \Km}\xspace}

  \def\Dbar    {\kern 0.2em\overline{\kern -0.2em \PD}{}\xspace}
\def\D       {\ensuremath{\PD}\xspace}

\def\Dz      {\ensuremath{\D^0}\xspace}
\def\Dzb     {\ensuremath{\Dbar^0}\xspace}
\def\DzDzb   {\ensuremath{\Dz {\kern -0.16em \Dzb}}\xspace}
\def\Dp      {\ensuremath{\D^+}\xspace}
\def\Dm      {\ensuremath{\D^-}\xspace}

\def\DpDm    {\ensuremath{\Dp {\kern -0.16em \Dm}}\xspace}

\def\Dstarp  {\ensuremath{\D^{*+}}\xspace}
\def\Dstarm  {\ensuremath{\D^{*-}}\xspace}

  \def\Bbar    {\kern 0.18em\overline{\kern -0.18em \PB}{}\xspace}

  \def\Y#1S{\ensuremath{\PUpsilon{(#1S)}}\xspace}% no space before {...}!

\def\to                 {\ensuremath{\rightarrow}\xspace}

\def\CP                {\ensuremath{C\!P}\xspace}

\def\FL       {\ensuremath{F_{\mathrm{L}}}\xspace}
\def\AT#1     {\ensuremath{A_{\mathrm{T}}^{#1}}\xspace}           % 2

\def\C#1      {\ensuremath{\mathcal{C}_{#1}}\xspace}                       % 9
\def\Cp#1     {\ensuremath{\mathcal{C}_{#1}^{'}}\xspace}                    % 7
\def\Ceff#1   {\ensuremath{\mathcal{C}_{#1}^{\mathrm{(eff)}}}\xspace}        % 9  
\def\Cpeff#1  {\ensuremath{\mathcal{C}_{#1}^{'\mathrm{(eff)}}}\xspace}       % 7
\def\Ope#1    {\ensuremath{\mathcal{O}_{#1}}\xspace}                       % 2
\def\Opep#1   {\ensuremath{\mathcal{O}_{#1}^{'}}\xspace}                    % 7

\newcommand{\tev}{\ensuremath{\mathrm{\,Te\kern -0.1em V}}\xspace}
\newcommand{\gev}{\ensuremath{\mathrm{\,Ge\kern -0.1em V}}\xspace}
\newcommand{\mev}{\ensuremath{\mathrm{\,Me\kern -0.1em V}}\xspace}
\newcommand{\kev}{\ensuremath{\mathrm{\,ke\kern -0.1em V}}\xspace}
\newcommand{\ev}{\ensuremath{\mathrm{\,e\kern -0.1em V}}\xspace}
\newcommand{\gevc}{\ensuremath{{\mathrm{\,Ge\kern -0.1em V\!/}c}}\xspace}
\newcommand{\mevc}{\ensuremath{{\mathrm{\,Me\kern -0.1em V\!/}c}}\xspace}
\newcommand{\gevcc}{\ensuremath{{\mathrm{\,Ge\kern -0.1em V\!/}c^2}}\xspace}
\newcommand{\gevgevcccc}{\ensuremath{{\mathrm{\,Ge\kern -0.1em V^2\!/}c^4}}\xspace}
\newcommand{\mevcc}{\ensuremath{{\mathrm{\,Me\kern -0.1em V\!/}c^2}}\xspace}

\def\mum  {\ensuremath{\,\upmu\rm m}\xspace}

\def\invfb   {\ensuremath{\mbox{\,fb}^{-1}}\xspace}

\def\gsim{{~\raise.15em\hbox{$>$}\kern-.85em
          \lower.35em\hbox{$\sim$}~}\xspace}
\def\lsim{{~\raise.15em\hbox{$<$}\kern-.85em
          \lower.35em\hbox{$\sim$}~}\xspace}

\def\tell1  {TELL1\xspace}
\def\ukl1   {UKL1\xspace}

\def\AP {\ensuremath{A_{\mathrm{P}}}\xspace}

\def\AD {\ensuremath{A_{\mathrm{D}}}\xspace}
\def\pis {\ensuremath{\pi_{\mathrm{s}}}\xspace}
\def\aindCP {\ensuremath{a^{\mathrm{ind}}_{\CP}}\xspace}
\def\adirCP {\ensuremath{a^{\mathrm{dir}}_{\CP}}\xspace}
\def\ARAW {\ensuremath{A_{\mathrm{raw}}}\xspace}
\def\ptFIXED {\mbox{$p_{\mathrm{T}}$}\xspace}

\usepackage{mciteplus}
\begin{document}
\title{Evidence for \boldmath \CP violation in time-integrated $\boldsymbol{D^0 \rightarrow h^-h^+}$ decay rates}
%\begin{center}
%The LHCb Collaboration
%\end{center}
\author{\begin{center}
R.~Aaij$^{23}$, 
C.~Abellan~Beteta$^{35,n}$, 
B.~Adeva$^{36}$, 
M.~Adinolfi$^{42}$, 
C.~Adrover$^{6}$, 
A.~Affolder$^{48}$, 
Z.~Ajaltouni$^{5}$, 
J.~Albrecht$^{37}$, 
F.~Alessio$^{37}$, 
M.~Alexander$^{47}$
% ...

{\footnotesize \it
% ...
$ ^{5}$Clermont Universit\'{e}, Universit\'{e} Blaise Pascal, CNRS/IN2P3, LPC, Clermont-Ferrand, France\\
$ ^{6}$CPPM, Aix-Marseille Universit\'{e}, CNRS/IN2P3, Marseille, France\\
% ...
$ ^{23}$Nikhef National Institute for Subatomic Physics, Amsterdam, The Netherlands\\
% ...
$ ^{35}$Universitat de Barcelona, Barcelona, Spain\\
$ ^{36}$Universidad de Santiago de Compostela, Santiago de Compostela, Spain\\
$ ^{37}$European Organization for Nuclear Research (CERN), Geneva, Switzerland\\
% ...
$ ^{42}$H.H. Wills Physics Laboratory, University of Bristol, Bristol, United Kingdom\\
% ...
$ ^{47}$School of Physics and Astronomy, University of Glasgow, Glasgow, United Kingdom\\
$ ^{48}$Oliver Lodge Laboratory, University of Liverpool, Liverpool, United Kingdom\\
% ...
$ ^{n}$LIFAELS, La Salle, Universitat Ramon Llull, Barcelona, Spain\\
% ...
}
\end{center}
}
\collaboration{The LHCb collaboration}
\begin{abstract}
%\begin{center}\emph{Submitted to Phys. Rev. Lett.}\end{center}
  \vspace{0.3cm}
  \noindent
   A search for time-integrated \CP violation in $\Dz \to h^-h^+$ ($h=K$, $\pi$) decays is presented using 0.62~\invfb of data collected by LHCb in 2011.
  The flavor of the charm meson is determined by the charge of the slow pion in the $\Dstarp \to \Dz \pip$ and $\Dstarm \to \Dzb \pim$ decay chains.
  The difference in \CP asymmetry between
  $\Dz \to K^- K^+$ and $\Dz \to \pi^- \pi^+$, $\Delta A_{\CP} \equiv  A_{\CP}(K^-K^+) \, - \, A_{\CP}(\pi^-\pi^+)$,  is measured to be
  $\left[ -0.82 \pm 0.21 (\mathrm{stat.}) \pm 0.11 (\mathrm{syst.}) \right]\%$.
  This differs from the hypothesis of \CP conservation by $3.5$ standard deviations.

\end{abstract}


\vspace*{-1cm}
\hspace{-9cm}
\mbox{\Large EUROPEAN ORGANIZATION FOR NUCLEAR RESEARCH (CERN)}

\vspace*{0.2cm}
\hspace*{-9cm}
\begin{tabular*}{16cm}{lc@{\extracolsep{\fill}}r}
\ifthenelse{\boolean{pdflatex}}% Logo format choice
{\vspace*{-2.7cm}\mbox{\!\!\!\includegraphics[width=.14\textwidth]{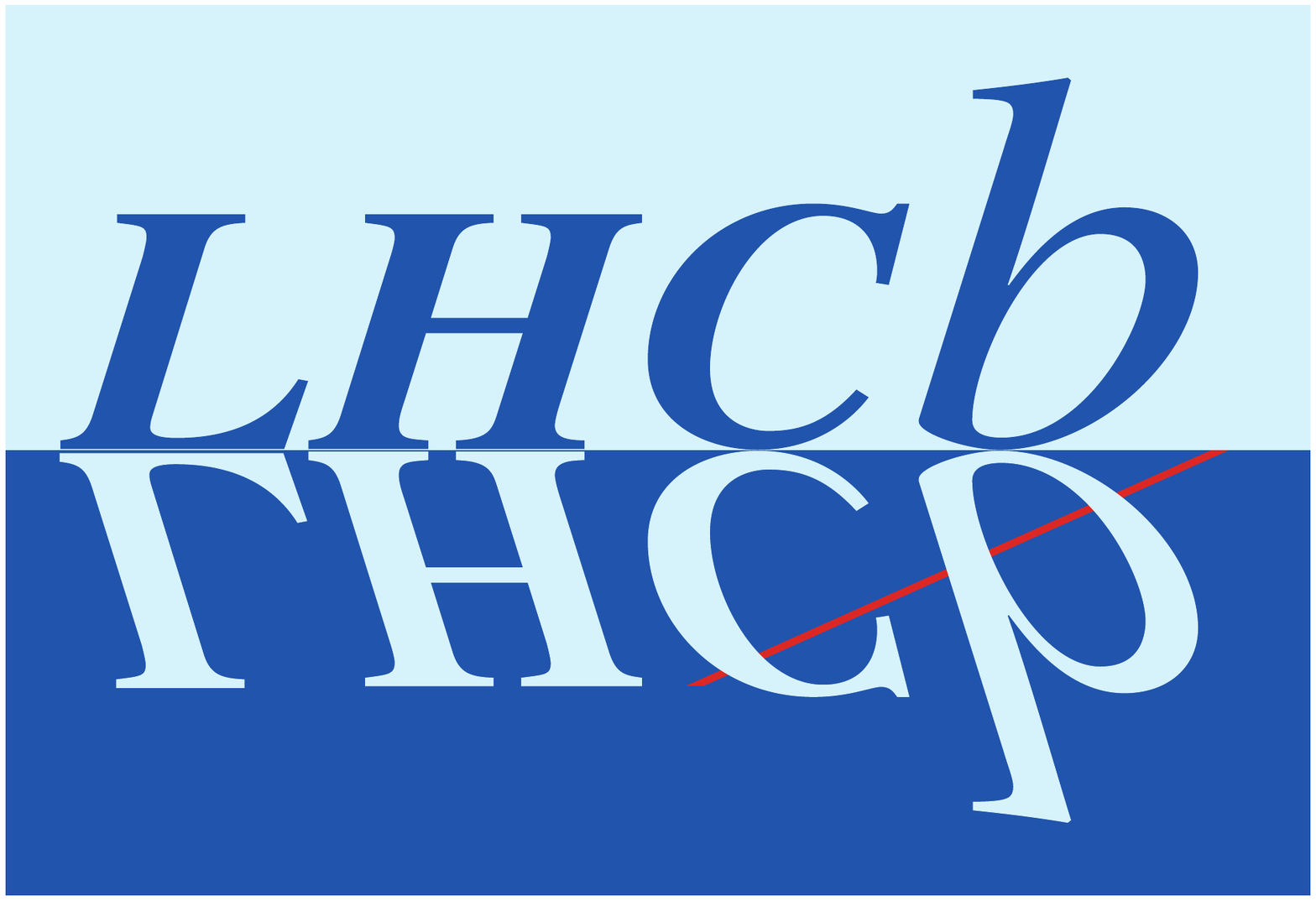}} & &}%
{\vspace*{-1.2cm}\mbox{\!\!\!\includegraphics[width=.12\textwidth]{lhcb-logo.eps}} & &}%
\\
 & & LHCb-PAPER-2011-023 \\
 & & CERN-PH-EP-2011-208 \\ % ID 
 & & \today \\ % Date - Can also hardwire e.g.: 23 March 2010
%  & & \\
% not in paper \hline
\end{tabular*}
\vspace*{1cm}

\maketitle

The charm sector is a promising place to  probe for the effects of physics
beyond the Standard Model (SM).
There has been a resurgence of interest in the past few years
since evidence for \Dz mixing was first
seen~\cite{bib:babar_mixing_moriond,bib:belle_mixing_moriond}. Mixing
is now well-established~\cite{bib:hfag} at a level which is consistent
with, but at the upper end of, SM expectations~\cite{falk_grossman_ligeti_nir_petrov}.
By contrast, no evidence for \CP violation in charm decays has yet been found.

The time-dependent \CP asymmetry $A_{\CP}(f;\,t)$ for $D^0$ decays to a CP eigenstate $f$ 
(with $f = \bar{f}$) is defined as
\begin{equation}
A_{\CP}(f;\,t) \equiv \frac{\Gamma(\Dz(t) \to f)-\Gamma(\Dzb(t) \to f)}{\Gamma(\Dz(t) \to f)+\Gamma(\Dzb(t) \to f)}, \label{eq:acpf}
\end{equation}
where $\Gamma$ is the decay rate for the process indicated.
In general $A_{\CP}(f;\,t)$ depends on $f$.
For $f= K^- K^+$ and $f= \pi^- \pi^+$, $A_{\CP}(f;\,t)$ can be expressed in terms of two contributions: 
  a direct component associated with \CP violation in the decay amplitudes, and
  an indirect component associated with \CP violation in the mixing or in the interference between mixing and decay.
In the limit of U-spin symmetry, the direct component is equal in magnitude and opposite in sign 
for $K^-K^+$ and $\pi^-\pi^+$, though the size of U-spin breaking effects remains to be
quantified precisely~\cite{bib:grossman_kagan_nir}.
The magnitudes of \CP asymmetries in decays to these final states are expected to be
small in the SM~\cite{bib:cicerone,bib:lenz,bib:grossman_kagan_nir,bib:petrov},
with predictions of up to $\mathcal{O}(10^{-3})$.
However, beyond the SM the rate of \CP violation
could be enhanced~\cite{bib:grossman_kagan_nir,bib:littlest_higgs}.

The asymmetry $A_{\CP}(f;\,t)$ may be written to first order as~\cite{bib:cdf_paper,bib:bigi_d2hh}
\begin{equation}
A_{\CP}(f;\,t) = \adirCP(f) \, + \, \frac { t }{\tau} \aindCP, \label{eq:acpphysicsth}
\end{equation}
where
$\adirCP(f)$ is the direct \CP asymmetry,
$\tau$ is the $D^0$ lifetime, and 
$\aindCP$ is the indirect \CP asymmetry.
To a good approximation this latter quantity is
universal~\cite{bib:grossman_kagan_nir,bib:kagan_sokoloff}.
The time-integrated asymmetry measured by an experiment, $A_{\CP}(f)$,
depends upon the time-acceptance of that experiment. It can be written as
\begin{equation}
A_{\CP}(f) = \adirCP(f) \, + \, \frac {\langle t \rangle}{\tau} \aindCP, \label{eq:acpphysics}
\end{equation}
where $\langle t \rangle$ is the average decay time in the reconstructed sample. Denoting by $\Delta$ the differences between quantities for $\Dz \to K^-K^+$ and $\Dz \to \pi^-\pi^+$ it is then possible to write
\begin{eqnarray}
\Delta A_{\CP} & \equiv & A_{\CP}(K^-K^+) \, - \, A_{\CP}(\pi^-\pi^+) \label{eq:acpfinal} \\
&  = & \left[ \adirCP(K^-K^+) \,-\, \adirCP(\pi^-\pi^+) \right] \, + \, \frac {\Delta \langle t \rangle}{\tau} \aindCP.\nonumber 
\end{eqnarray}
In the limit that  $\Delta \langle t \rangle$ vanishes,
$\Delta A_{\CP}$ is equal to the difference in the direct \CP asymmetry
between the two decays.
However, if the time-acceptance is different for the $\Km\Kp$ and $\pim\pip$ final
states, a contribution from indirect \CP violation remains.

The most precise measurements to date of the time-integrated \CP
asymmetries in $\Dz \to K^- K^+$ and $\Dz \to \pi^- \pi^+$ were made
by the CDF, BaBar, and Belle
collaborations~\cite{bib:cdf_paper,bib:babar_paper2008,bib:belle_paper2008}.
The Heavy Flavor Averaging Group (HFAG) has combined time-integrated
and time-dependent measurements of \CP asymmetries, taking account of
the different decay time acceptances, to obtain world average
values for the indirect \CP asymmetry of $a_{\CP}^{\mathrm{ind}} = (-0.03 \pm 0.23)\%$ and
the difference in direct \CP asymmetry between the final states of $\Delta a_{\CP}^{\mathrm{dir}} = (-0.42 \pm 0.27)\%$~\cite{bib:hfag}.

In this Letter, we present a measurement
of the difference in time-integrated \CP asymmetries between $D^0 \to K^-K^+$
and $D^0 \to \pi^-\pi^+$, performed with 0.62~\invfb of data collected 
at LHCb between March and June 2011. 
The flavor of the initial state (\Dz or \Dzb) is tagged by
requiring a $\Dstarp \to \Dz \pis^+$ decay,
with the flavor determined by the charge of the slow pion ($\pis^+$).
The inclusion of charge-conjugate modes is implied throughout,
except in the definition of asymmetries.

The raw asymmetry for tagged \Dz decays to a final state $f$
is given by $\ARAW(f)$, defined as
\begin{equation}
\ARAW(f) \equiv \frac{N(D^{*+} \to D^0( f)\pi_s^+) \, - \, N(D^{*-} \to \Dzb (f)\pi_s^-)}
                                            {N(D^{*+} \to D^0( f)\pi_s^+) \, + \, N(D^{*-} \to \Dzb (f)\pi_s^-)},
\label{def:astarrawdef}
\end{equation}
where $N(X)$ refers to the number of reconstructed events of decay $X$
after background subtraction. 

To first order the raw asymmetries may be written as a sum of four components,
due to physics and detector effects:
\begin{equation}
\ARAW(f) = A_{\CP}(f) \, + \, \AD(f) \, + \, \AD(\pis^+) \, + \, \AP(D^{*+}).
\label{def:arawstarcomponents}
\end{equation}
Here, $\AD(f)$ is the asymmetry in selecting the $D^0$ decay
into the final state $f$,  
$\AD(\pis^+)$ is the asymmetry in selecting the slow pion
from the $D^{*+}$ decay chain, and 
$\AP(D^{*+})$ is the production asymmetry
for $D^{*+}$ mesons.
The asymmetries $\AD$ and $\AP$ are defined in
the same fashion as \ARAW.
The first-order expansion is valid since the individual asymmetries are small.

For a two-body decay of a spin-0 particle to a self-conjugate
final state there can be no $D^0$ detection asymmetry,
i.e. $\AD(K^-K^+) = \AD(\pi^-\pi^+) = 0.$  
Moreover, $\AD(\pis^+)$ and $\AP(D^{*+})$ are independent of $f$ and thus in
the first-order expansion of equation 5 those terms cancel in the difference
$\ARAW(K^-K^+) \, - \, \ARAW(\pi^-\pi^+)$, resulting in
\begin{equation}
\Delta A_{\CP} = \ARAW (K^-K^+) \, - \, \ARAW (\pi^-\pi^+).\label{eq:adefequals}
\end{equation}
To minimize second-order effects that are related to the slightly different kinematic
properties of the two decay modes and that do not cancel in $\Delta A_{\CP}$,
the analysis is performed in bins of the relevant kinematic variables, as discussed later.

The LHCb detector is a forward spectrometer covering the pseudorapidity range
$2 < \eta < 5$, and is described in detail in Ref.~\cite{LHCb}.
The Ring Imaging Cherenkov (RICH) detectors are of particular
importance to this analysis, providing kaon-pion
discrimination for the full range of track momenta used.
The nominal downstream beam direction is aligned with the $+z$ axis, and
the field direction in the LHCb dipole is such that charged 
particles are deflected in the horizontal ($xz$) plane.  The field polarity was changed
several times during data taking: about 60\% of the data were taken
with the down polarity and 40\% with the other.

Selections are applied to provide samples of $D^{*+}\to D^0 \pis^+$ 
candidates, with $D^0 \to K^-K^+$ or $\pi^-\pi^+$.
Events are required to pass both
hardware and software trigger levels. A loose $D^0$ selection is applied
in the final state of the software trigger, and
in the offline analysis only candidates that are accepted by this
trigger algorithm are considered.
Both the trigger and offline selections impose a variety of requirements on
kinematics and decay time to isolate the decays of interest, 
including requirements
  on the track fit quality,
  on the \Dz and \Dstarp vertex fit quality,
  on the transverse momentum ($\ptFIXED > 2$~GeV$/c$) and decay time ($ct > 100 \, \mum$) of the \Dz candidate,
  on the angle between the \Dz momentum in the lab frame and its
    daughter momenta in the \Dz rest frame  ($|\cos \theta| < 0.9$),
  that the \Dz trajectory points back to a primary vertex,
  and that the \Dz daughter tracks do not.
In addition, the offline analysis exploits the capabilities of the RICH
system to distinguish between pions and kaons when reconstructing the \Dz meson,
with no tracks appearing as both pion and kaon candidates.

A fiducial region is implemented by imposing the requirement
that the slow pion lies within the central part of the detector acceptance.
This is necessary because the magnetic field bends pions of one charge
to the left and those of the other charge to the right. For soft tracks at large angles in the $xz$ plane this
implies that one charge is much more likely to remain within the 300~mrad horizontal detector
acceptance, thus making  $\AD(\pis^+)$ large.
Although this asymmetry is formally independent of the \Dz decay mode,
it breaks the assumption that the raw asymmetries are small and
it carries a risk of second-order systematic effects if the ratio of 
efficiencies of $\Dz \to \Km\Kp$ and $\Dz \to \pim\pip$ varies in the affected region.
The fiducial requirements therefore exclude edge regions 
in the slow pion $(p_x, p)$ plane. Similarly, a small
region of phase space in which one charge of slow pion is more likely to be
swept into the beampipe region in the downstream tracking stations,
and hence has reduced efficiency, is also excluded. After the implementation of the fiducial requirements about 70\% of the events are retained.

\begin{figure}
  \begin{center}
   \includegraphics[width=0.49\textwidth]{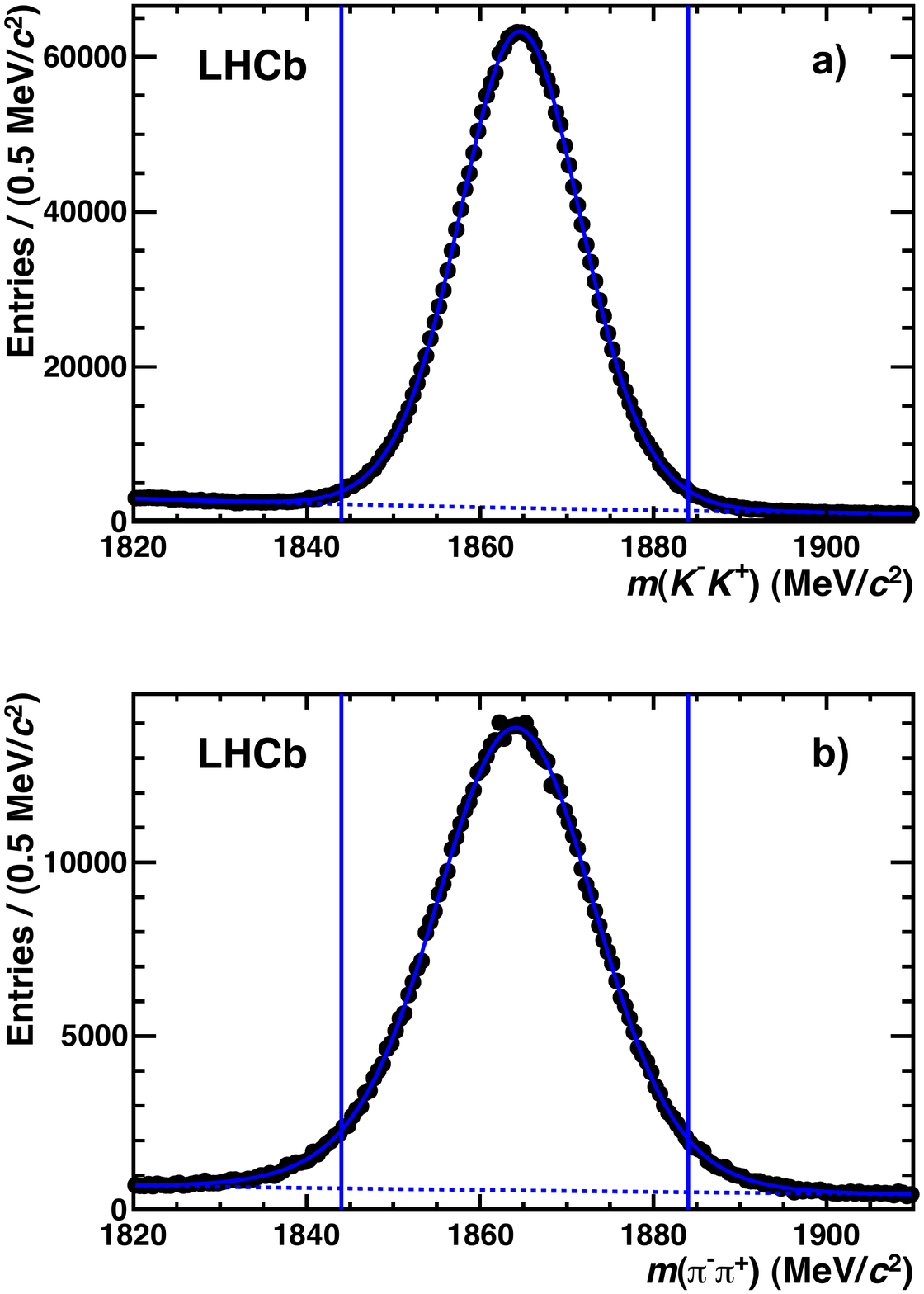}
   \vspace{-1cm}
   \end{center}
  \caption{
    Fits to the (a) $m(\Km\Kp)$ and (b) $m(\pim\pip)$ spectra of \Dstarp candidates passing
    the selection and satisfying $0<\delta m<15$~MeV$/c^2$.
    The dashed line corresponds to the background component in the fit, and
    the vertical lines indicate the signal window of 1844--1884~MeV$/c^2$.
 }
  \label{fig:massTagged}
\end{figure}

\begin{figure}
  \begin{center}
   \includegraphics[width=0.49\textwidth]{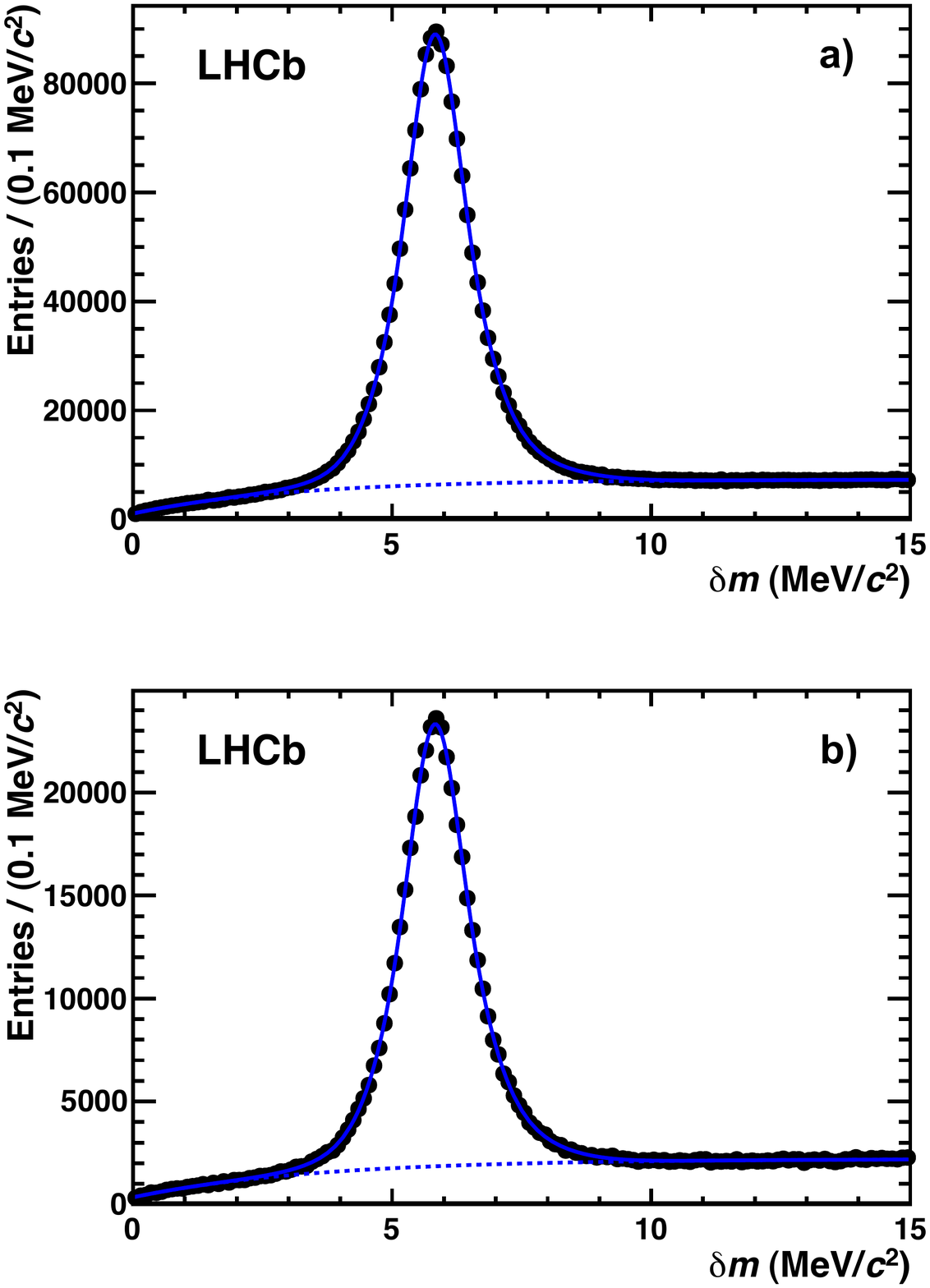}
   \vspace{-1cm}
   \end{center}
  \caption{
    Fits to the $\delta m$ spectra, where the $D^0$ is reconstructed
    in the final states (a) $K^- K^+$ and (b) $\pi^- \pi^+$, 
    with mass lying in the window of 1844--1884~MeV$/c^2$.
    The dashed line corresponds to the background component in the fit.
 }
  \label{fig:deltaMassTagged}
\end{figure}

The invariant mass spectra of selected $K^-K^+$ and $\pi^-\pi^+$ pairs are shown in Fig.~\ref{fig:massTagged}. The half-width at half-maximum of the signal lineshape is
8.6~MeV$/c^2$ for $\Km\Kp$ and 11.2~MeV$/c^2$ for $\pim\pip$, where the difference is
due to the kinematics of the decays and has no relevance for the subsequent analysis.
The mass difference ($\delta m$) spectra of selected candidates, where $\delta m \equiv m(h^- h^+ \pis^+) - m(h^- h^+) - m(\pi^+)$
for $h=K,\pi$, are shown in Fig.~\ref{fig:deltaMassTagged}.
Candidates are required to lie inside
a wide $\delta m$ window of 0--15~MeV$/c^2$, and in
Fig.~\ref{fig:deltaMassTagged} and for all subsequent results candidates
are in addition required to lie in a mass signal window of 1844--1884~MeV$/c^2$.
The \Dstarp signal yields are approximately
$1.44 \times 10^6$ in the $K^-K^+$ sample,
and $0.38 \times 10^6$ in the $\pi^-\pi^+$ sample.
Charm from $b$-hadron decays is strongly suppressed by the
requirement that the \Dz originate from a primary vertex, 
and accounts for only 3\% of the total yield.
Of the events that contain at least one \Dstarp candidate,
12\% contain more than one candidate; this is expected due
to background soft pions from the primary vertex and all candidates are accepted.
The background-subtracted average decay time of $D^0$ candidates passing 
the selection is measured for each final state, and the fractional difference
$\Delta \langle t \rangle / \tau$ is obtained.
Systematic uncertainties on this quantity are assigned for
  the uncertainty on the world average \Dz lifetime $\tau$ $(0.04\%)$,
  charm from $b$-hadron decays $(0.18\%$), and
  the background-subtraction procedure $(0.04\%)$.
Combining the systematic uncertainties in quadrature, we obtain
$\Delta \langle t \rangle / \tau = \left[ 9.83 \pm 0.22 (\mathrm{stat.}) \pm 0.19 (\mathrm{syst.}) \right] \%$.
The $\pi^-\pi^+$ and $K^-K^+$ average decay time is $\langle t \rangle=\left( 0.8539 \pm 0.0005 \right )$~ps, where the error is statistical only. 

Fits are performed on the samples in order to determine
$\ARAW(K^-K^+)$ and $\ARAW(\pi^-\pi^+)$.
The production and detection 
asymmetries can vary with \ptFIXED and pseudorapidity $\eta$, and so can the 
detection efficiency of the two different $D^0$ decays, in particular 
through the effects of the particle identification requirements. 
The analysis is performed in 54 kinematic bins defined
by the \ptFIXED and $\eta$ of the $D^{*+}$ candidates,
the momentum of the slow pion, and the sign of $p_x$ of the
slow pion at the \Dstarp vertex.
The events are further partitioned in two ways.
First, the data are divided between the two dipole magnet polarities.
Second, the first 60\% of data are processed separately from
the remainder, with the division aligned with a
break in data taking due to an LHC technical stop.
In total, 216 statistically independent measurements are considered for each decay mode.

In each bin, one-dimensional unbinned maximum likelihood fits to the $\delta m$ spectra are performed.
The signal is described as the sum of two Gaussian functions with a common
mean $\mu$ but different widths $\sigma_i$, convolved with a function $B(\delta m; s) = \Theta(\delta m) \, \delta m^s$
taking account of the asymmetric shape of the measured $\delta m$
distribution. Here, $s \simeq -0.975$ is a shape parameter fixed to the value determined from the global fits shown in Fig. \ref{fig:deltaMassTagged},
$\Theta$ is the Heaviside step function, 
and the convolution runs over $\delta m$.
The background is described by an empirical function of the form
$1 - e^{-(\delta m - \delta m_0)/\alpha}$,
where $\delta m_0$ and $\alpha$ are free parameters describing the threshold and shape of the
function, respectively. 
The \Dstarp and \Dstarm samples in a given bin are fitted simultaneously
and share all shape parameters, except for a charge-dependent offset in the central
value $\mu$ and an overall scale factor in the mass resolution. The raw asymmetry in the
signal yields is extracted directly from this simultaneous fit.
No fit parameters are shared between the 216 subsamples of data,
nor between the $\Km\Kp$ and $\pim\pip$ final states.

The fits do not distinguish between
the signal and backgrounds that peak in
$\delta m$. Such backgrounds can arise from \Dstarp decays in which the correct slow pion is found but the \Dz is partially mis-reconstructed. These backgrounds are suppressed by the use of tight
particle identification requirements and a narrow $D^0$ mass window. From studies of
the $D^0$ mass sidebands (1820--1840 and 1890--1910~MeV$/c^2$), this contamination is found to
be approximately 1\% of the signal yield and to have small raw asymmetry
(consistent with zero asymmetry difference between the $\Km\Kp$ and $\pim\pip$ final states).
Its effect on the measurement is estimated in
an ensemble of simulated experiments and found to be negligible; a systematic uncertainty
is assigned below based on the statistical precision of the estimate.

A value of $\Delta A_{\CP}$ is determined in each measurement bin
as the difference between $\ARAW(K^-K^+)$ and $\ARAW(\pi^-\pi^+)$.
Testing these 216 measurements for mutual consistency, we obtain
$\chi^2/\mathrm{ndf} = 211/215$ ($\chi^2$ probability of 56\%).
A weighted average is performed to yield the result $\Delta A_{\CP} =  (-0.82 \pm 0.21 )\%$,
where the uncertainty is statistical only.

\begin{figure}
  \begin{center}
    \includegraphics[width=0.49\textwidth]{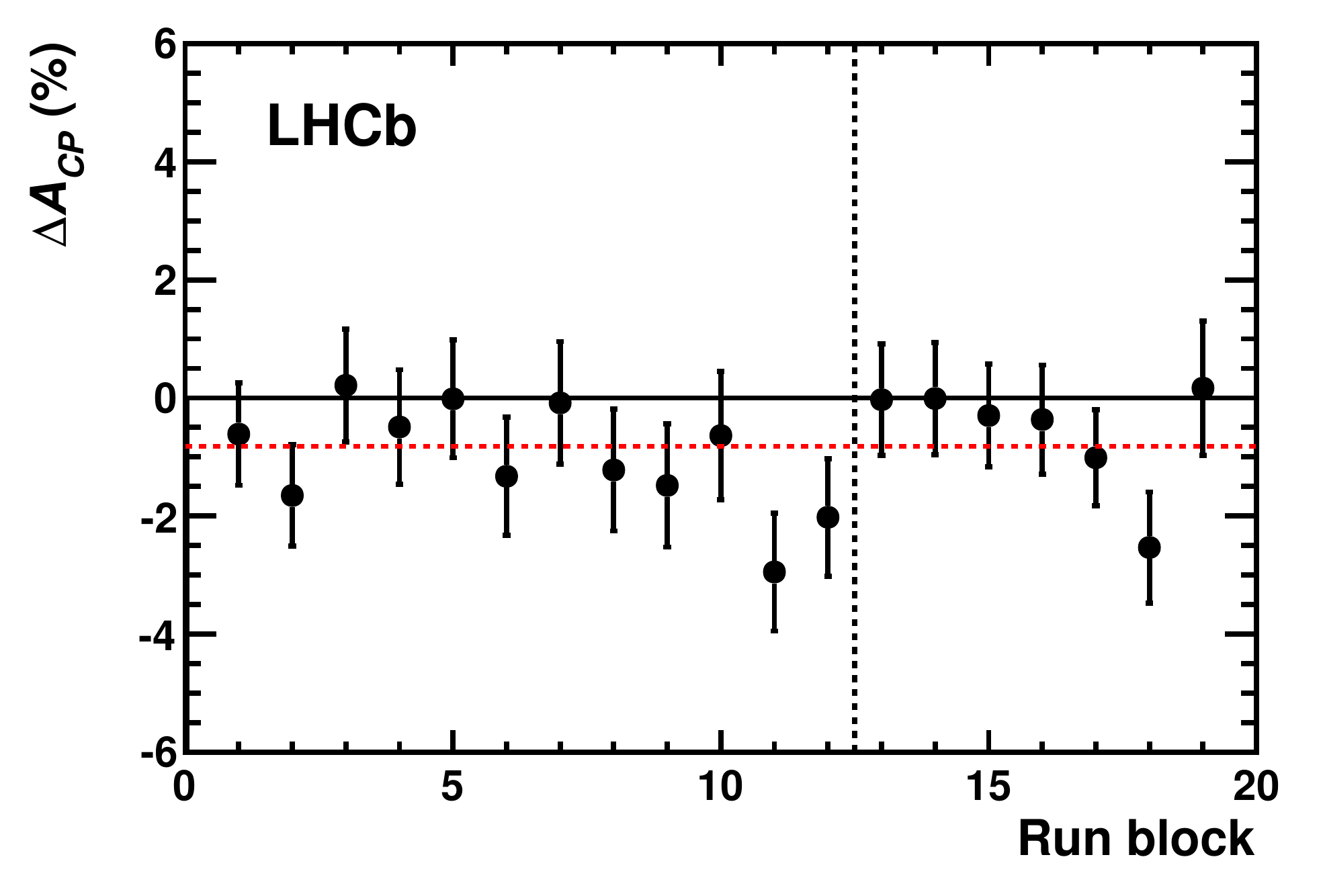}
   \end{center}
  \caption{
    Time-dependence of the measurement. The data are divided into
    19 disjoint, contiguous, time-ordered blocks and the value of
    $\Delta A_{\CP}$ measured in each block. 
    The horizontal red dashed line shows the result for the combined sample.
    The vertical dashed line indicates the technical stop
    referred to in Table~\ref{tab:subsamples}.
  }
  \label{fig:time}
\end{figure}

Numerous robustness checks are made.
The value of $\Delta A_{\CP}$ is studied as a function of the time at which the
data were taken (Fig.~\ref{fig:time}) and found to be consistent with
a constant value ($\chi^2$ probability of 57\%). The measurement is repeated
with progressively more restrictive RICH particle identification requirements,
finding values of $(-0.88 \pm 0.26)\%$ and $(-1.03 \pm 0.31)\%$; 
both of these values are consistent with the baseline result when correlations are taken into account. 
Table~\ref{tab:subsamples} lists $\Delta A_{\CP}$ for eight disjoint
subsamples of data split according to magnet polarity, the sign of $p_x$
of the slow pion, and whether the data were taken before or after
the technical stop. The $\chi^2$ probability for consistency among the
subsamples is 45\%. The significances of the differences between data
taken before and after the technical stop, between the magnet polarities,
and between $p_x>0$ and $p_x<0$ are $0.4$, $0.6$, and
$0.7$ standard deviations, respectively.
Other checks include 
applying electron and muon vetoes to the slow pion and to the \Dz daughters, 
use of different kinematic binnings,
validation of the size of the statistical uncertainties with Monte Carlo pseudo-experiments,
tightening of kinematic requirements, 
testing for variation of the result with the multiplicity of tracks and of primary vertices in the event, 
use of other signal and background parameterizations in the fit, and imposing a full set of common shape parameters between $D^{*+}$ and $D^{*-}$ candidates.
Potential biases due to the inclusive hardware trigger selection are investigated
with the subsample of data in which one of the signal final-state tracks is directly
responsible for the hardware trigger decision. 
In all cases good stability is observed.
For several of these checks, a reduced number of kinematic bins are used
for simplicity. No systematic dependence of $\Delta A_{\CP}$ is observed with respect to the kinematic variables.

\begin{table}
  \caption{
      Values of $\Delta A_{\CP}$ measured in subsamples of the data,
      and the $\chi^2/\mathrm{ndf}$ and corresponding $\chi^2$ probabilities 
      for internal consistency among the 27 bins in each subsample.
      The data are divided before and after a technical stop (TS), by magnet
      polarity (up, down), and by the sign of $p_x$ for the slow pion
      (left, right).
      The consistency among the eight subsamples is
      $\chi^2/\mathrm{ndf}=6.8/7$ (45\%).
  }
  \begin{center}
    \begin{ruledtabular}
      \begin{tabular}{lcc}
        Subsample & $\Delta A_{\CP}~[\%]$ & $\chi^2/\mathrm{ndf}$  \\
        \colrule
        Pre-TS,   up,   left  & $-1.22 \pm 0.59$ & $13/26~(98\%)$ \\
        Pre-TS,   up,   right & $-1.43 \pm 0.59$ & $27/26~(39\%)$ \\
        Pre-TS,   down, left  & $-0.59 \pm 0.52$ & $19/26~(84\%)$ \\
        Pre-TS,   down, right & $-0.51 \pm 0.52$ & $29/26~(30\%)$ \\
        Post-TS,  up,   left  & $-0.79 \pm 0.90$ & $26/26~(44\%)$ \\
        Post-TS,  up,   right & $+0.42 \pm 0.93$ & $21/26~(77\%)$ \\
        Post-TS,  down, left  & $-0.24 \pm 0.56$ & $34/26~(15\%)$ \\
        Post-TS,  down, right & $-1.59 \pm 0.57$ & $35/26~(12\%)$ \\
        \hline
        All data & $-0.82 \pm 0.21$ & $211/215~(56\%)$ \\
      \end{tabular}
    \end{ruledtabular}
  \end{center}
\label{tab:subsamples}
\end{table}

Systematic uncertainties are assigned 
  by: loosening the fiducial requirement on the slow pion;
  assessing the effect of potential peaking backgrounds in Monte Carlo pseudo-experiments;
  repeating the analysis with the asymmetry extracted through sideband subtraction in $\delta m$ instead of a fit;
  removing all candidates but one (chosen at random) in events with multiple candidates;
  and comparing with the result obtained without kinematic binning.
In each case the full value of the change in result is taken as
the systematic uncertainty. These uncertainties are listed in
Table~\ref{tab:systematics:summary_acp_kk_pp}.  The sum in quadrature is $0.11\%$.
Combining statistical and systematic uncertainties in quadrature,
this result is consistent at the $1\sigma$ level with the current
HFAG world average~\cite{bib:hfag}.

\begin{table}
  \caption{
    Summary of absolute systematic uncertainties for $\Delta A_{\CP}$.
  }
  \begin{center}
    \begin{ruledtabular}
      \begin{tabular}{lc}
        Source & Uncertainty \\ 
        \colrule
        Fiducial requirement & 0.01\% \\
        Peaking background asymmetry & 0.04\% \\
        Fit procedure & 0.08\% \\
        Multiple candidates & 0.06\% \\
        Kinematic binning & 0.02\% \\
        \hline
        Total & 0.11\% \\
      \end{tabular}
    \end{ruledtabular}
  \end{center}
  \label{tab:systematics:summary_acp_kk_pp}
\end{table}

In conclusion,
the time-integrated difference in \CP asymmetry between $D^0 \to K^-K^+$ and $D^0 \to \pi^-\pi^+$ decays has been measured to be
\begin{displaymath}
  \Delta A_{\CP} = \left[ -0.82 \pm 0.21 (\mathrm{stat.}) \pm 0.11 (\mathrm{syst.}) \right]\%
\end{displaymath}
with 0.62~\invfb of 2011 data.
Given the dependence of $\Delta A_{\CP}$ on the direct and indirect \CP asymmetries, shown in Eq. (\ref{eq:acpfinal}), and the measured value
$\Delta \langle t \rangle / \tau = \left[ 9.83 \pm 0.22 (\mathrm{stat.}) \pm 0.19 (\mathrm{syst.}) \right] \%$, the contribution
from indirect \CP violation is suppressed and $\Delta A_{\CP}$ is primarily
sensitive to direct \CP violation.
Dividing the central value by the sum in quadrature of the
statistical and systematic uncertainties,
the significance of the measured deviation from zero is $3.5\sigma$. This is the first evidence for \CP violation in the charm sector.
To establish whether this result is consistent with the SM will require the analysis of more data, as well as improved theoretical understanding.

\section*{Acknowledgements}

\noindent We express our gratitude to our colleagues in the CERN accelerator
departments for the excellent performance of the LHC. We thank the
technical and administrative staff at CERN and at the LHCb institutes,
and acknowledge support from the National Agencies: CAPES, CNPq,
FAPERJ and FINEP (Brazil); CERN; NSFC (China); CNRS/IN2P3 (France);
BMBF, DFG, HGF and MPG (Germany); SFI (Ireland); INFN (Italy); FOM and
NWO (The Netherlands); SCSR (Poland); ANCS (Romania); MinES of Russia and
Rosatom (Russia); MICINN, XuntaGal and GENCAT (Spain); SNSF and SER
(Switzerland); NAS Ukraine (Ukraine); STFC (United Kingdom); NSF
(USA). We also acknowledge the support received from the ERC under FP7
and the Region Auvergne.

\bibliographystyle{LHCb}
\bibliography{main-lhcb}

\end{document}